\ifdefined\pdfcreationdate\else
  \edef\pdfcreationdate{\string D:20261107101435+00'00'}
\fi
\documentclass{ceurart}
\usepackage{tabularx}
\usepackage{xcolor}
\definecolor{QuoteBlue}{HTML}{6473A6}
\newcolumntype{Y}{>{\raggedright\arraybackslash}X}
\ExplSyntaxOn
\NewDocumentCommand{\placeholder}{m}{
  \group_begin:
  \setlength{\fboxsep}{0pt}
  \seq_set_split:Nnn \l_tmpa_seq { ~ } { #1 }
  \seq_map_indexed_inline:Nn \l_tmpa_seq {
    \int_compare:nNnT { ##1 } > { 1 } { \space }
    \colorbox{yellow!35}{\strut ##2}
  }
  \group_end:
}
\ExplSyntaxOff
\begin{document}
\copyrightyear{2027}
\copyrightclause{Copyright for this paper by its authors. Use permitted under Creative Commons License Attribution 4.0 International (CC BY 4.0).}
\conference{HIMAS 2027: Human-Centered Interactive Multimodal AI Systems, co-located with ACM IUI 2027, February 2027, Helsinki, Finland}
\title{Exploring multimodal AI--XR interactions for software development: From expert perspectives to semantic code outlines}
\author[1]{Ilya Zakharov}[email=ilia.zaharov@jetbrains.com]
\cormark[1]
\author[1]{Alexey Markin}
\author[1]{Konstantin Ostroukhov}
\author[1]{Alexey Perminov}
\author[2]{Roham Koohestani}
\author[2]{Ekaterina Koshchenko}
\author[1]{Agnia Sergeyuk}
\address[1]{JetBrains Research, Belgrade, Serbia}
\address[2]{JetBrains Research, Amsterdam, Netherlands}
\begin{abstract}
AI tools can generate code quickly, but developers still need to understand and check the results. We explore how the enriched multimodal context derived from the extended reality (XR) workspace might help developers manage this information. We first interviewed 13 experts in AI, XR, and related fields to identify opportunities and design challenges (Study 1). These interviews informed two prototypes explored with 12 professional developers (Study 2): four tried a version where the code faded to become invisible after it was looked at, and eight tried a version that replaced viewed code sections with short descriptions of their purpose. In the second prototype, head direction helped determine when to offer a change, while manual input let developers apply it and reopen the original code. A separate open-ended demonstration with seven software-engineering researchers provided complementary feedback. Participants’ accounts suggested that descriptions could provide an overview of unfamiliar code and serve as reminders when revisiting it. However, difficulty comparing descriptions with source code, headset limitations, and controls that deviates from regular IDE routines could introduce additional effort. These exploratory findings identify requirements for attention support: make summaries easy to recognize and access, preserve straightforward inspection of the original code, and retain familiar navigation and editing practices. The study contributes to human-centered multimodal systems by showing how the coordination of input channels and AI-generated representations must account for users’ attention, verification needs, and established ways of working.

\end{abstract}
\maketitle

\section{Introduction}
Artificial Intelligence (AI) is changing software production and interaction with development tools. Coding agents can undertake larger changes, while developers remain responsible for the design and quality of the resulting software \citep{Huang2026Control}. Spatial representation and rich multimodal inputs in extended reality (XR) open a design space for human--AI interaction in software development. Alongside head, gaze, gesture, and speech tracking, emerging sensing options include wearable electroencephalography (EEG) \citep{Kosnacova2026EEG,PiEEG2026XR}, photoplethysmography (PPG) for cognitive-load monitoring in XR \citep{Alshehhi2026PPG}, and compact motion and contextual sensors for headsets and glasses \citep{BoschSensortecXR}. AI can help interpret this context \citep{Hirzle2023}. This creates a human-centered multimodal design problem: how should an interface use that context to manage visible detail while preserving developers’ ability to inspect, compare, and recover source code \citep{Pfeuffer2024,Khan2026, Sergeyuk2026}?.

The growing volume of AI-generated code motivates exploring the design space of AI-XR interaction for software engineering. Producing more code does not provide equivalent support for understanding or maintaining it. Early Copilot users struggled to understand, edit, and debug generated snippets \citep{Vaithilingam2022}; developer forum discussions document review burdens from low-quality AI-generated code and related contributions \citep{Baltes2026}. Alakmeh et al.'s eye-tracking study found that about half of AI code suggestions were not looked at, and more than three-quarters of those inspected were unused, identifying repeated micro-interruptions as an interaction cost \citep{Alakmeh2026}. Anthropic reported approval of roughly 93\% of Claude Code permission prompts and described approval fatigue \citep{Anthropic2026Containment}. Meanwhile, professional developers actively sought to steer coding agents and retain control over design and implementation \citep{Huang2026Control}. These accounts motivate research on navigating and assessing generated work, without implying that all generated code has these problems.

We investigate interactions where the growing cognitive demands due to the amount of AI-generated code in software engineering meet multimodal AI and XR. We first interviewed 13 experts in AI, XR, and related fields to identify opportunities and design challenges (Study 1). These interviews informed two prototypes explored with 12 professional developers: four tried a  pilot version that faded parts of the code, and eight tried a version that replaced selected code sections with short descriptions of their purpose (Study 2). Study 1 explores expert perspectives on technology creation, including software development. Study 2 examines one resulting direction through iterative prototypes for exploring unfamiliar generated code. A separate free demonstration with computer-science/software-engineering (CS/SE) researchers supplies complementary feedback. We ask:

\textbf{Overarching RQ:} What interaction opportunities and design requirements arise when multimodal AI and XR are brought into software development?

\textbf{RQ1 --- Expert interviews:} What needs, interaction opportunities, and constraints should guide multimodal AI-XR support for technology creation?

\textbf{RQ2 --- Prototype work:} How can adaptive code representations support developers' attention during XR code exploration, and what interaction requirements emerge from their use?

We contribute an expert-derived design space, a formative investigation connecting it to developer experience, and requirements for coordinating representation, activation, and access to source.

\section{Background and related work}
\textbf{Interaction demands in AI-assisted development.} AI assistance changes what developers communicate and inspect. Snippet generation required understanding, correction, and debugging \citep{Vaithilingam2022}; agentic development adds task delegation, steering, and deciding what to automate \citep{Huang2026Control}. Developers also seek project context, multi-file support, configurable assistance, and infrastructure integration \citep{Sergeyuk2026}, broadening the design problem beyond a correct next suggestion.

Alakmeh et al. examine visual engagement and interruptions around inline suggestions \citep{Alakmeh2026}, whereas Baltes et al. analyze accounts of review friction and maintenance costs \citep{Baltes2026}. Neither establishes XR as a remedy; both motivate interfaces for inspecting and acting on generated artifacts within a larger workflow.

\textbf{Connecting views and levels of code.} Program comprehension involves moving between higher-level hypotheses and implementation details, revisiting source, and following program relationships \citep{vonMayrhauser1995}. Initial orientation and checking exact behavior require different information. Developers' agent-control strategies also depend on expertise and quality judgments \citep{Huang2026Control}. Interfaces must therefore support changing information needs.

Requests for context and integration \citep{Sergeyuk2026} concern connected work, beyond the amount displayed. We explore which familiar operations people expect to retain and where new representations or channels could extend them.

\textbf{Implicit context and explicit action.} Context sensing and explicit commands provide different evidence of intention. Adaptive and user-adaptable interfaces distribute control differently \citep{Findlater2004}; mixed-initiative principles and human--AI guidelines connect inference with timing, correction, and intervention \citep{Horvitz1999,Amershi2019}. Gaze + Pinch separates visual targeting from manual selection \citep{Pfeuffer2017,Pfeuffer2024}, while gaze-and-speech interaction combines situated context with spoken requests \citep{Khan2026}.

A sensed orientation need not be an instruction to transform content, and repeated confirmation need not guarantee meaningful supervision \citep{Anthropic2026Containment}. The design question is what context should inform assistance, which decisions require an explicit action, and how people can inspect or override the result. We use this distinction when examining the interaction developed after Study 1.

\section{Research approach and methods}
We used a sequential exploratory approach (Figure 1). Study 1 addressed RQ1 by mapping expert perspectives on multimodal AI-XR support for technology creation. We selected attention support for code exploration from that design space and developed RQ2 for Study 2. The second study was formative and consisted of building and testing two prototypes: the pilot informed a design change, and the settled outline phase examined useful roles, interaction difficulties, and requirements for refinement. The expert interviews addressed technology creation broadly, while the prototypes focused on software development in particular. The cohorts contribute different kinds of evidence and are not pooled into one sample.  
Both studies followed JetBrains institutional ethical standards and the values and guidelines of the ICC/ESOMAR International Code \citep{ICCESOMAR2024}.

\begin{figure}[htbp]
\centering
\includegraphics[width=0.90\linewidth]{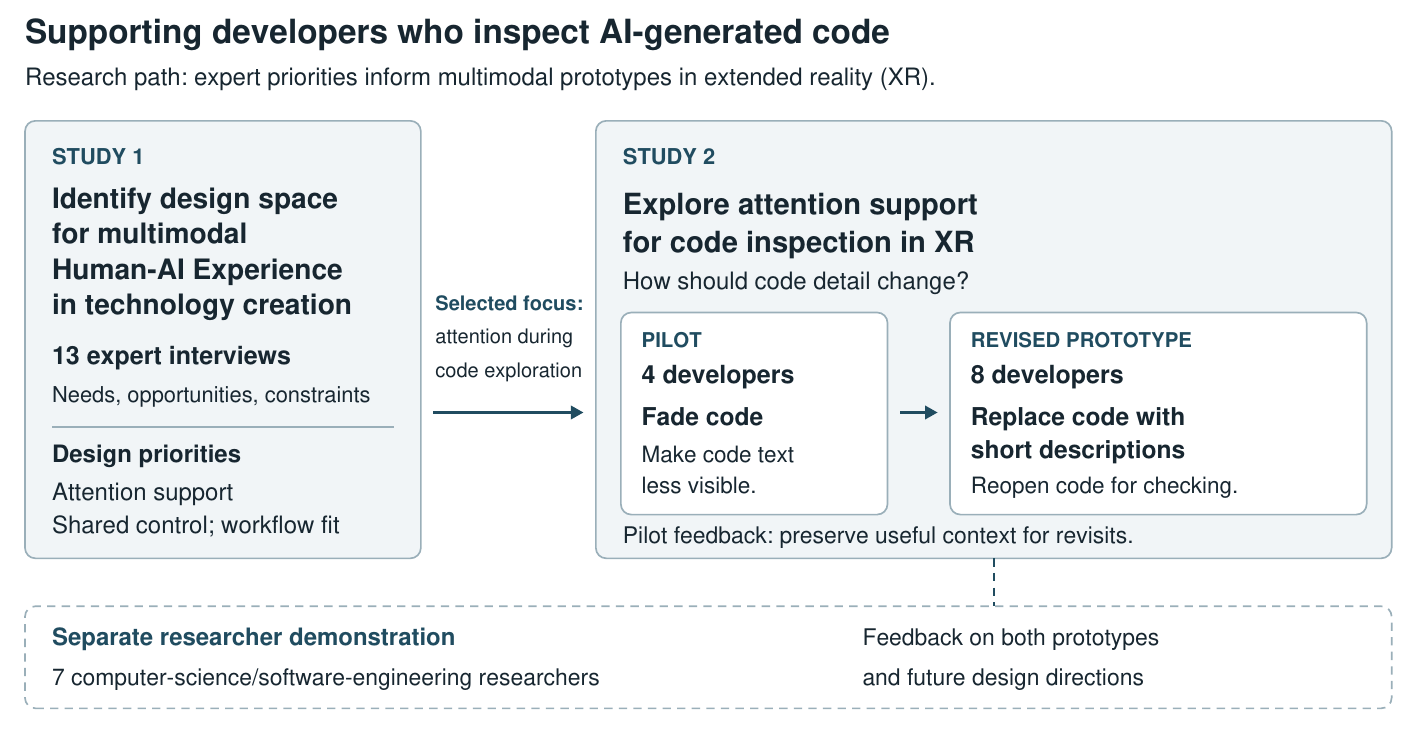}
\caption{Research sequence. Expert interviews identified design priorities for multimodal AI in extended reality. The team selected attention support during code exploration, first piloting text fading and then examining reversible code descriptions. A separate researcher demonstration provided complementary feedback on both prototypes.}
\label{fig:research-flow}
\end{figure}
\subsection{Study 1: Expert interviews}
We interviewed 13 experts in AI, XR, multimodal interaction, and related research/product leadership. Purposive recruitment considered academic research and professional recognition, and industry leadership, working groups, patents, and open-source contributions. Two pilot interviews refined the guide before 11 further interviews. Sessions lasted approximately 45--90 minutes across four months, ending in February 2026.

The guide covered AI-XR experience, prototypes, modalities, technology-creation needs, imagined tools, adoption, and risks. Braun and Clarke's six-phase thematic analysis guided coding and theme development \citep{Braun2006}. Two authors independently coded the first eight interviews and reconciled differences; a second researcher reviewed subsequent coding. Analysis was revisited during recruitment. Expert identifiers are P1--P13.

\subsection{Study 2: Participants and procedure}
Study 2 involved 12 internally recruited JetBrains developers from July to September 2026. After four participants for the pilot version with code fading (\textit{diminution prototype}), we developed an  updated prototype; there were eight participants for the updated version (replacing code with short descriptions; \textit{natural language (NL) outline prototype}). We focused on the comprehension of AI-generated code as the task for the prototype as the most representative part of today's software engineering work. Using PICO 4 Ultra headset and WebXR, participants explored unfamiliar AI-generated Java code, verbalized their experience, and completed a retrospective semi-structured interview about useful and disruptive moments, control, confidence in transformations, workflow fit, and adoption. Post-session questionnaires about the adaptive functionality supplied descriptive context: the System Usability Scale (SUS) \citep{Brooke1996}, short User Experience Questionnaire (UEQ-S) \citep{Schrepp2017}, and Raw NASA Task Load Index (NASA TLX) for perceived workload. Sessions lasted 40--60 minutes.

The pilot varied Java task material and interaction with static and diminished presentations to inform design selection. All eight outline sessions then used a common format to examine the resulting interaction. Identifiers are D1--D4 and N1--N8, respectively.

Of the 12 participants, eight reported 6--10 years of professional coding experience and four reported 11--15 years. All resided in Belgrade, Serbia, held intermediate or senior engineering roles, and reported limited or no XR experience. Java experience ranged from one to seven years; one respondent described two to three years on secondary projects more than five years earlier. Recruitment used internal JetBrains employee channels; participants received no monetary compensation.

A separate open-ended demonstration of both prototypes on PICO 4 Ultra involved seven CS/SE researchers, R1--R7 with more than 3 years of research experience, no overlap with the developer cohorts. Outline-related remarks and workflow discussion supplied complementary design evidence. Median prototype viewing time during free demonstration was 10 minutes.

\subsection{Prototype and stimuli}
WebXR prototypes on PICO 4 Ultra displayed source alongside its generation prompt. A camera-forward ray used head orientation for coarse exposure/targeting context; manual input supported selection and confirmation.

\begin{figure}[htbp]
\centering
\includegraphics[width=0.80\linewidth]{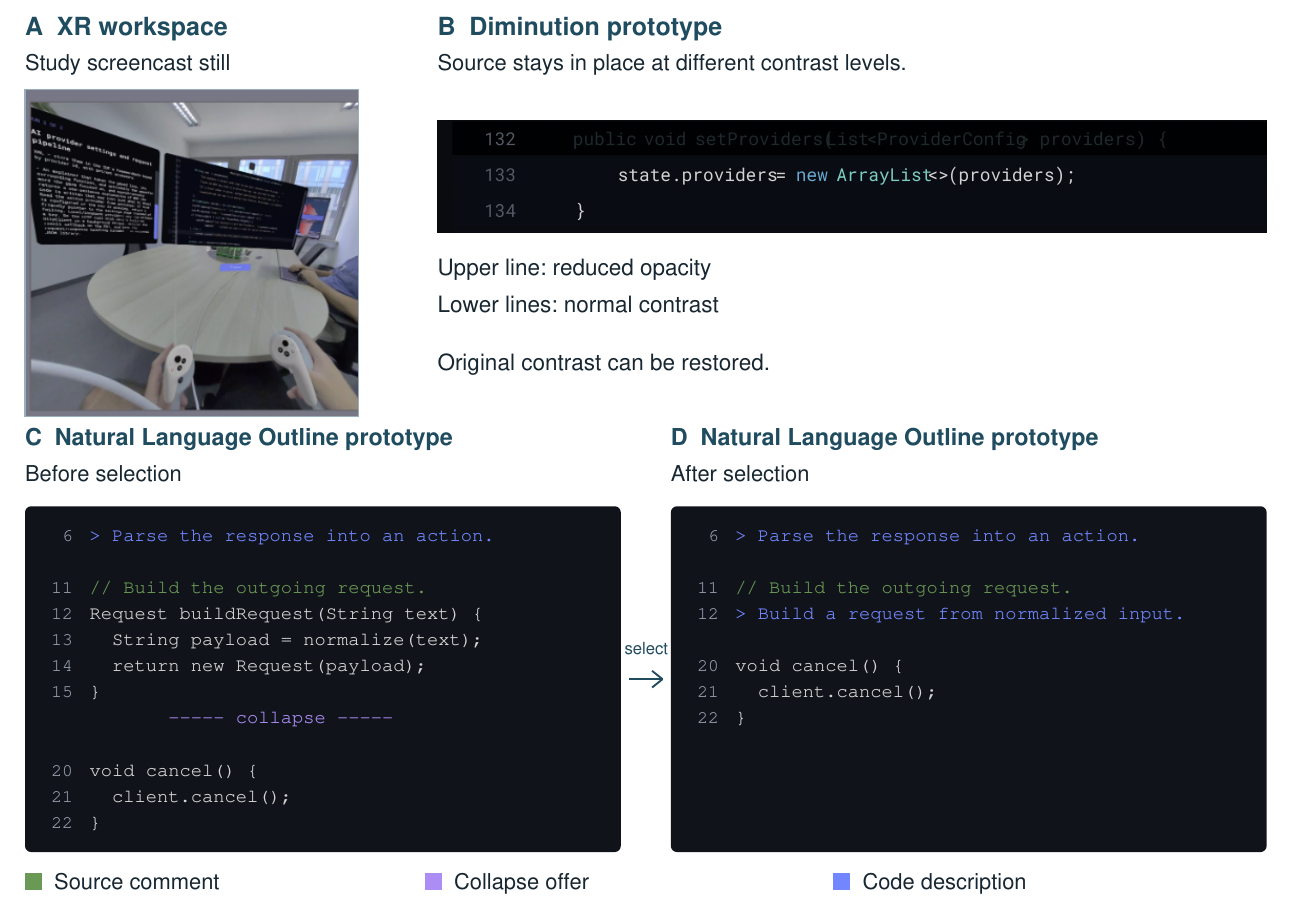}
\caption{Prototype views for XR code exploration. (A) A study screencast showing spatial text and code panels and handheld controllers. (B) Diminution Prototype: the upper code line has reduced opacity; lower lines retain normal contrast. The source stays in place and its contrast can be restored. (C--D) Natural Language Outline prototype schematics with illustrative Java code. (C) Expanded code and an existing outline coexist; head-directed exposure makes a purple collapse offer available. (D) Selecting to collapse replaces the region with a blue description of its purpose; the selection is reversible. Green denotes an ordinary source comment.}
\label{fig:prototype-overview}

\label{fig:diminution-view}
\label{fig:prototype-views}
\end{figure}

The \textbf{Diminution prototype} reduced text opacity after an exposure rule was met, with restoration available. Here \emph{diminution} means attenuating displayed content so that it becomes less visible~(Figure~\ref{fig:diminution-view}). The \textbf{Natural Language Outline prototype} instead replaced a selected code region with a short description of its purpose \citep{Shi2025}. Source was initially visible; accumulated head-directed exposure made an inline collapse offer available. Manual selection replaced that region with its outline, and selecting the outline reopened the source. Expanded regions and outlines could coexist. Sensing thus offered an action that the developer chose whether to apply (Figure~\ref{fig:prototype-views}).

Stimuli comprised AI-generated Java components of an IntelliJ plugin: input routing, calibration, AI requests, and a gaze engine, each about 200 lines long. They included generation prompts, semantic-region boundaries, and descriptions at several detail levels.

\subsection{Prototypes evaluation approach}
Study 2's main analysis covered eight retrospective interviews and screen-recording transcripts of spoken accounts during exploration, without systematic frame-by-frame behavioral coding. Interview and screen accounts from one person remain one case. We used codebook thematic analysis, as in Study 1. For Study 2, Sol 6.1 with Extra High reasoning assisted transcript organization, rereading, initial coding, and candidate thematic interpretations. The first author reviewed the preparation and updated the codes and themes; two co-authors then checked the analysis and the resulting codes. Interviews and screen recordings were transcribed locally with Whisper large-v3-turbo. The analysis retains 177 meaningful passages and 20 operational codes in five provisional themes. Multiple codes can apply to one passage; counts aid retrieval rather than estimate prevalence.

Pilot analysis supplied design rationale, and the separate open-ended demonstration for CS/SE researchers reused the framework while allowing additional design propositions.

\section{Study 1: The interaction design space}
The AI-XR experts interview analysis produced 156 codes in five themes: professional creation, contextual AI, new interaction paradigms, readiness and adoption friction, and ethics and human control (Table~\ref{tab:codebook}). These describe the design space, not participant frequencies. Detailed findings are described as a separate preprint \citep{Zakharov2026Space}; three connections guided the prototype study.

\begin{table*}[!htbp]
  \caption{Overview of the final Design Space codebook: five overarching themes, their guiding questions, and second-level subgroups (156 codes in total).}
  \label{tab:codebook}
  \footnotesize
  \setlength{\tabcolsep}{4pt}
  \begin{tabularx}{\textwidth}{@{}>{\raggedright\arraybackslash}p{0.30\textwidth}>{\raggedright\arraybackslash}Xc@{}}
    \toprule
    \textbf{Theme and guiding question} & \textbf{Subgroups (number of codes)} & \textbf{Total} \\
    \midrule
    AI-XR for creation, collaboration, and professional work\par
      \emph{How does AI-XR change how people make things?}
      & Spatial workspaces and portable offices (6); Spatial visualization and embodied engineering (5); AI-assisted prototyping and design (9); Collaboration and representation (2); Knowledge work and the changing creator role (6); Training and immersive professional domains (3)
      & 31 \\
    \addlinespace
    AI as contextual intelligence layer\par
      \emph{Can AI make XR environments understand and adapt to you?}
      & World models and scene understanding (5); Human signals and behavioral context (4); Intent inference and mediation (7); Personalization, attention, and proactivity (8); Situated augmentation and environment control (9)
      & 33 \\
    \addlinespace
    Readiness, adoption, and ecosystem friction\par
      \emph{What is actually blocking XR-AI from becoming mainstream?}
      & Hardware and infrastructure bottlenecks (7); Sensing, precision, and system performance (3); Wearability, ergonomics, and public acceptability (9); Adoption value and workflow friction (9); Paradigm and ecosystem immaturity (10)
      & 38 \\
    \addlinespace
    New interaction paradigms\par
      \emph{How do humans communicate intent to AI-XR systems?}
      & Interaction paradigm shift (7); Multimodal intent orchestration (8); Input channels and wearable controls (7); Embodied presence and multimodal feedback (8); Interaction breakdowns (4)
      & 34 \\
    \addlinespace
    Ethics, control, and the human position\par
      \emph{As AI-XR systems grow more capable and ambient, what do humans lose or risk?}
      & Privacy, surveillance, and security (5); Regulation, governance, and social norms (5); Human agency and dependence (3); Transparency, accountability, and ownership (3); Manipulation, commercialization, and wellbeing (4)
      & 20 \\
    \bottomrule
  \end{tabularx}
\end{table*}

First, experts saw opportunities to connect fragmented work through spatial workspaces, contextual assistance, and movement between ideas and implementation. Finite attention constrained contextual AI: assistance should filter and organize information rather than continually add to it. This motivated our attention-management focus.

Second, experts argued that combining inputs required an understandable model of shared control. As \textbf{P2} put it in the expert interview, \textcolor{QuoteBlue}{``We have to learn how to share control between a human and a machine.''}

Third, adoption and ethics themes identified constraints: headset comfort, sensing precision, workflow friction, privacy, and agency. Our resulting prototype priorities were to simplify information, retain orientation, make transformations reversible, and preserve developers' inspection and control. These are design responses to the interviews, rather than features validated by the experts.

\section{Interim discussion: Selecting a prototype direction}
Expert findings identified opportunities for spatial organization, contextual assistance, richer expression of intent, and movement between ideas and implementation. Across these, assistance must fit finite human attention and preserve understandable control. We selected attention support during code exploration, connecting experts' concern about continually adding information to inspecting unfamiliar generated material. Developers move between an overview, exact implementation, and related source \citep{vonMayrhauser1995}. RQ2 therefore asks how representations can make detail easier to choose while preserving access for revisits.

Our three design commitments were to vary salience or detail, retain orientation and recoverable source, and separate contextual sensing from deliberate transformation. These were team decisions informed by the interviews. Head direction supplied practical context on PICO 4 Ultra, while manual action expressed whether to apply a change; these choices do not establish expert preference for these channels or a universal need for manual confirmation.

\section{Study 2: Formative prototype evaluation}

Study 2 progressed from attenuating exposed text to retaining a short description of a hidden region's purpose. Its formative aim was to inform subsequent design and evaluation, rather than compare the iterations as matched experimental conditions. 

\subsection{Pilot: From diminution and fading to semantic outlines}
The four-person pilot lowered code salience after registered exposure. D1 encountered little fading. D2 preferred the static example, reporting uncertain restoration and lost context. D3 reported more deliberate reading but questioned line-by-line processing for familiar patterns. D4 described changes drawing attention back to the interface and reducing readability.

Plausible uses included marking progress in linear checks and hiding completed work. However, inspected code could still be needed for cross-reference. The team therefore moved from uniform line-level fading and diminution to semantic outlines that could preserve a region's purpose while reversible expansion supported revisits.

\subsection{Natural Language Outline themes}

The eight-case analysis of the main NL prototype identified five themes. They connect the useful roles of semantic abstraction with the work required to obtain, check, and act on it. Interview analysis is presented below; supporting questionnaires are reported briefly afterward.

\begin{table}[htbp]
\caption{Evidence from all eight outline participants. The table preserves conditional and conflicting accounts rather than ranking acceptance. All eight contributed interviews; seven also have screen-recording transcripts with concurrent accounts. N6 contributes interview evidence only.}
\label{tab:3}
\footnotesize
\setlength{\tabcolsep}{4pt}
\begin{tabularx}{\linewidth}{@{}>{\raggedright\arraybackslash}p{0.10\linewidth}YY@{}}
\toprule
\textbf{Outline case} & \textbf{Recorded interaction or account} & \textbf{Qualification for the design claim} \\
\midrule
N1 & Initially treated the outline as an ordinary comment; recognized the abstraction after explanation. & Outlining must be recognizable before its semantic value can be assessed. \\
N2 & Used a chosen preview and returned to source; preferred source initially visible. & Preview value does not imply a preference for a fully collapsed starting view. \\
N3 & Valued logical chunk sizes; wanted summary and implementation available together. & Useful abstraction and acceptable source access are separate design decisions. \\
N4 & Appreciated folding after guidance; proposed documentation preview and questioned checking effort. & Coached discovery and proposed uses differ from unaided uptake. \\
N5 & Initial interest gave way to control-seeking, scope uncertainty, and restoring source. & Assistance can add coordination work when availability is hard to anticipate. \\
N6 & Valued semantic reminders for revisits; wanted comments, editing, and familiar input. & Outline value extends beyond XR; focus/memory agreement followed a moderator prompt. \\
N7 & Wanted co-present source/summary, a preview of the affected lines, a persistent folding arrow, and multi-file views. & Concurrent accounts ground these requests; the proposed remedies were not evaluated and inspection was too shallow to assess cognitive benefit. \\
N8 & Accepted the summary idea conditionally; requested predictable folding and habitual language-server navigation. & Rarely used ordinary folding; offer-seeking, eye strain, and shallow inspection qualified personal fit. \\
\bottomrule
\end{tabularx}
\end{table}

\textbf{Entry points into unfamiliar code.} N2 collapsed a complex-looking block for an overview, then returned to the implementation, while preferring source initially visible. Choosing a preview did not imply wanting everything initially collapsed. N4 proposed documentation summaries to select detail for inspection. N6 saw descriptions as landmarks when revisiting referenced functions and wanted semantic folding in an ordinary IDE. N5 sought descriptions of difficult regions before the exposure rule made controls available. These accounts suggest three roles: choosing where to inspect, compressing inspected detail, and retaining landmarks for revisits. Preview access should be distinct from an offer to hide detail after exposure.

Participants' existing strategies differed. N4 described folding everything in IDEs before selectively opening relevant regions, whereas N8 normally kept source open and rarely used ordinary folding. N3 liked the logical chunk sizes while opposing an exclusive source/summary representation. These contrasts support configurable starting views and selective use of outlines, rather than a single assumed reading strategy.

\textbf{Predictable scope and availability.} N5 captured difficulty anticipating transformations in the retrospective interview: \textcolor{QuoteBlue}{``I don't understand what will appear.''}
N5's initial interest shifted toward searching for controls and reopening source to restore consistent detail. N7 reported offers disappearing before activation and extra scrolling to find collapse actions after reopening blocks, proposing a persistent gutter arrow beside the line numbers. N8 also described repeated effort to elicit offers; in the interview, the participant clarified that summarization itself was not distracting and requested predictable, scope-based behavior. Interest in the representation differed from the effort of obtaining it. N2 found outlining unobtrusive despite unavailable controls; N4 appreciated it after guidance; N5 proposed persistent function/class controls. Stable manual access, visible eligibility, and inexpensive reversal could preserve this value without requiring users to monitor assistance; these refinements remain untested.

\textbf{Recognizable and verifiable summaries.} N1 initially interpreted an outline as a comment and discussed its semantic role after explanation. N5 wanted descriptions distinct from source; N8 wanted the collapse control to resemble a button. Transformed content and available actions both need recognizable visual roles.

During code exploration, N3 described the cost of mutually exclusive views: collapsing \textcolor{QuoteBlue}{``hides the implementation itself, so you can't cross-reference it''}.
N7 independently preferred simultaneous source and summary and requested the affected start and end lines before collapse. Scope is thus an inspection requirement: developers need to know which implementation a description represents before hiding it. N2 questioned annotation provenance; N4 anticipated checking effort but found one prompted spot-check approximately consistent. These accounts support meaningful boundaries, clear provenance, and inexpensive verification without establishing summary inaccuracy. N7 found short descriptions insufficiently informative but had not inspected the code deeply enough to judge cognitive benefit. Requests for richer explanations or simultaneous source identify information needs, without demonstrating improved understanding.

\textbf{Continuity with developer practices.} Participants distinguished outlining from the XR workspace. N2 valued postural flexibility and N4 appreciated surrounding screens; N3 and N5 questioned the headset's added value. N7 proposed simultaneous views of different files for comparison.

Familiar operations were central. N1 requested find-usages and documentation. N6 stressed that review requires comments or edits, questioning how to preserve them while holding controllers. N8 described a two-editor workflow with adjacent documents and project structure. In the concurrent account, N8 identified habitual language-server navigation: clicking through to a module's source or interface. Its absence made the static surface harder to use; no language-server integration was evaluated. N6 also wanted outlines in an ordinary IDE. Semantic and spatial support should preserve dependable navigation, input, and editing while accommodating personal habits. Interest in NL outlines in general and in the headset are separate judgments.

\textbf{A stable visual workspace.} Reported blur, small targets, wrong-panel scrolling, display movement, and tiredness could compete with semantic compression. N7 described distracting word-by-word reflow and paragraph movement in a prompt panel, concerning the surrounding interface rather than an established effect of outlining. Stable text and reliable targets are part of usable assistance.

N2 appreciated comfortable posture and adjustable placement. N1 requested a plain background and reported easier concentration after covering the passthrough cameras, a case-specific account rather than a cohort-wide benefit. Layout stability, fit, target acquisition, and environmental control merit evaluation separately from the semantic representation.

\subsection{Questionnaire descriptions}

Questionnaires contextualize the themes. SUS averaged 59.6 (SD 16.5; median 65; range 35--80), indicating varied perceived usability. All eight respondents expected most people to learn the adaptive functionality quickly, although confidence and willingness to use it frequently were mixed. UEQ-S pragmatic and hedonic quality averaged 0.64 (SD 1.54) and 0.79 (SD 1.06), respectively, both near neutral. Raw TLX averaged 28.6 (SD 6.2) out of 100 for the adaptive NL outline prototype. These descriptions suggest a low anticipated learning barrier alongside varied practical fit; they do not establish workload reduction.

\subsection{Complementary researcher demonstration}

The separate demonstration supplied additional support and insights into the prototypes. R1 and R2 appraised outlining favorably; R3 accepted selective folding of boilerplate while retaining important code for repeated inspection. R6 tentatively accepted the function but questioned gaze's role. R4 and R7 criticized the interaction and suggested initially outlined functions, expandable for detail. R6 proposed progressively revealing explanation and source during code review or merge-conflict inspection. R1 proposed explanations during source reading alongside a collapsed NL view. R4 compared the interaction with familiar integrated development environment (IDE) folding controls beside functions in large files, supporting learned operations and clear semantic value. According to the demonstration, agency depends on understandable, controllable behavior and an appropriate way to express intent. Discussion also proposed linked definitions, code graphs, product-to-code navigation, and persistent context cards. These remain prospective requirements. Table~\ref{tab:2} integrates design conclusions across phases. 

\begin{table}[htbp]
\caption{Evidence-to-design joint display. Case examples identify support and qualifications, not prevalence or effect size. Researcher proposals remain distinct from observed interaction with the prototype.}
\label{tab:2}
\footnotesize
\setlength{\tabcolsep}{4pt}
\begin{tabularx}{\linewidth}{@{}>{\raggedright\arraybackslash}p{0.20\linewidth}YY@{}}
\toprule
\textbf{Design issue} & \textbf{Evidence and qualification} & \textbf{Implication for the next design} \\
\midrule
Preserve useful context & Pilot loss-of-context accounts; N3 wanted source and description together; R3 would retain important code. & Support selective compression, reopening, and simultaneous checking where needed. \\
Make abstraction available & N2 used a chosen preview; N5 sought unavailable descriptions; R6/R7 proposed earlier summaries. & Offer preview independently of exposure-gated compression. \\
Make transformations legible & N1 did not initially recognize outlining; N5 questioned scope and availability; R4 reported a target mismatch. & Distinguish summaries visually and expose the region and action rule. \\
Let channels serve intent & The prototype combines head targeting and confirmation; R2 preferred hands-free action; R6 questioned gaze's role. & Separate context sensing from activation; provide an appropriate override or fallback. \\
Give spatial presentation a role & Posture and surrounding screens were appreciated; developer and researcher cases requested relational views. & Keep useful context accessible and evaluate the value of spatial organization explicitly. \\
\bottomrule
\end{tabularx}
\end{table}

\section{Discussion across the studies}

\subsection{From expert priorities to prototype requirements}

Developers saw potential in short AI-generated code descriptions \cite{Shi2025}'for overviews and reminders, but reported various requirements to reduce the amount of frictions during this interaction mode.  These findings translate the experts’ priorities of attention support and shared control into requirements for human-centered multimodal systems: predictable controls that go in line with the UI learned in the previous coding environments, descriptions linked to source, and reversible changes. Head direction provides context, while manual selection lets developers apply or reverse changes. Researcher requests for hands-free and proactive assistance also motivate flexible ways to express intent. Quanitative attention and comprehension benefits remain to be tested in the future research.

\subsection{Representation, activation, spatial organization, and familiarity}

Three requirements interact: \textbf{representation} determines what replaces hidden detail; \textbf{activation}, when and how changes are requested; and \textbf{spatial organization}, where related source and context remain accessible. Alongside recognition, visible scope, and verification, the reversibility is also a critical demand: a distinct outline appearance, range preview, stable controls, immediate expansion, and simultaneous summary/source views. Similar interaction approach was recently introduced in ConceptEVA, where people could explore and customize information needs \citep{Zhang2023}. Our findings extend previous result concerning the relationship between description, source, and the switching action, connecting the mixed initiative interaction \citep{Horvitz1999,Amershi2019} to the of AI-generated work \citep{Huang2026Control}.

Simplifying code can complicate the surrounding interaction. The next design should establish dependable manual outlining before examining contextual offers. Text reflow, target acquisition, and environmental distraction need separate attention so their costs are not attributed to semantic compression alone. Eye tracking could refine targeting and speech or gestures alter activation, without determining what information should remain. Requests for hands-free and proactive assistance caution against equating agency with manual effort. Systems should expose their policies, permit intervention, and explain context collection and its beneficiaries.

A future integration with IDE or language-server services could preserve semantic operations and keyboard access while keeping related definitions, explanations, and review context visible together. Contextual input could locate material and explicit action control transformations. This functionality remains unevaluated.

\subsection{Implications for multimodal human-AI systems}

Multimodal input can make intentions easier to express by combining information that different channels convey well. Bolt's classic ``Put-That-There'' concept illustrates how speech specifies an action while pointing identifies its target \citep{Bolt1980}. For human-AI interaction, this suggests a way to connect requests to the visible workspace without describing every object in words. Combining channels can also resolve ambiguities in individual inputs. Multimodality offers flexibility: users can change channels with the task rather than combine them for every action \citep{Oviatt1999}. However, additional inputs introduce ambiguity about which signals belong together and whether behavior expresses a command. They also create learning demands: Norman argues that bodily gestures do not automatically make interfaces understandable \citep{Norman2010}. Our study illustrates this tension because head orientation supplied context without establishing when developers wanted less detail. Human-centered multimodality therefore requires a balance between richer contextual information for AI and the effort people expend making themselves understood.

\subsection{Scope and limitations}

Study 1 captures expert perspectives rather than a representative forecast. Study 2's small, internally recruited cohorts had short prototype encounters shaped by guidance and headset constraints. The eight outline sessions used a common format, but pilot iteration cannot establish comparative efficacy of fading and outlining. Head direction provides just a coarse orientation context. Future research is needed to verify the results with more fine-grained eye tracking data. The free demonstration supplies complementary design feedback rather than a controlled replication of the prototype evaluation.

\section{Conclusion}

Across two formative studies, we connected expert priorities for multimodal human--AI interaction with developers' experiences of code presentation in XR. Participants saw value in natural-language overviews, while difficulties accessing outlines exposed a gap between sensed behavior and intended action. These findings frame attention support as a coordination problem involving what information is shown, when it changes, and how people influence those changes. Multimodal inputs offer contextual information but also introduce uncertainty about intent. Our contribution is an empirical account of these opportunities and tensions, informing the design of human-centered multimodal assistance that fits people's ongoing work.

\section*{Declaration on Generative AI}
GPT 6.1 Sol with Extra High reasoning assisted transcription, preliminary coding, and refinement of Study 2 interviews, as well as manuscript editing, and LaTeX/figure layouts. GPT-5.6 Sol assisted generating both the Java components as the stimuli and the NL outlines. The authors take responsibility for the final analysis and publication content.

\bibliography{references}
\end{document}